\documentclass[preprint,aps,12pt,showpacs,nofootinbib,tightenlines]{revtex4}
\usepackage{amsmath}
\usepackage{amssymb}
\usepackage{epsfig}
\usepackage{graphicx}
\begin{document}
\newcommand{\beq}{\begin{eqnarray}}
\newcommand{\eeq}{\end{eqnarray}}

\newcommand{\bxsga}{B\to X_s \gamma}
\newcommand{\brbxsga}{{\cal B}(B\to X_s \gamma)}
\newcommand{\bzbzb}{ B_d^0 - \hat{B}_d^0 }

\newcommand{\bsga}{  b\to s \gamma}
\newcommand{\bdga}{  b\to d \gamma}
\newcommand{\bvga}{  B\to V \gamma }
\newcommand{\bksga}{ B\to K^* \gamma}
\newcommand{\brhoga}{B\to \rho \gamma}

\newcommand{\brbkz}{{\cal B}(B\to \overline{K}^{*0} \gamma)}
\newcommand{\brbkm}{{\cal B}(B\to K^{*-} \gamma)}
\newcommand{\brbrm}{{\cal B}(B\to \rho^- \gamma)}
\newcommand{\brbrz}{{\cal B}(B\to \rho^0 \gamma)}

\newcommand{\calb}{ {\cal B}}
\newcommand{\acp}{ {\cal A}_{CP}}
\newcommand{\oas}{ {\cal O} (\alpha_s)}

\newcommand{\mt}{m_t}
\newcommand{\mw}{M_W}
\newcommand{\mhp}{M_{H}}
\newcommand{\muw}{\mu_W}
\newcommand{\mub}{\mu_b}
\newcommand{\dmd}{\Delta M_{B_d} }
\newcommand{\ltt}{\lambda_{tt} }
\newcommand{\lbb}{\lambda_{bb} }
\newcommand{\rhob}{\hat{\rho} }
\newcommand{\etab}{\hat{\eta} }

\newcommand{\smallsm}{{\scriptscriptstyle SM}}
\newcommand{\smallyy}{{\scriptscriptstyle YY}}
\newcommand{\smallxy}{{\scriptscriptstyle XY}}
\newcommand{\smallnp}{{\scriptscriptstyle NP}}

\newcommand{\tab}[1]{Table \ref{#1}}
\newcommand{\fig}[1]{Fig.\ref{#1}}
\newcommand{\real}{{\rm Re}\,}
\newcommand{\im}{{\rm Im}\,}
\newcommand{\non}{\nonumber\\ }

\title{Signal of New Physics and Test of Isospin-SU(3) Relations and CP Violation in Charmless B Decays}
\author{ Yue-Liang Wu$^a$}
\email{ylwu@itp.ac.cn} 
%
\author{Yu-Feng Zhou$^b$ }
\email{zhou@post.kek.jp}
\author{Ci Zhuang$^a$ }
\email{zhuangc@itp.ac.cn}
\affiliation{(a)\ Institute of
theoretical physics, Chinese Academy of Science, Beijing 100080, China \\
(b)\ Theory Group, KEK, Tsukuba 305-0801, Japan}
\date{\today}

\begin{abstract}
A model-independent analytical analysis for charmless B decays is
presented. It is demonstrated that the CP-averaging branching ratio
difference $\Delta R = R_c - R_n$ in $B\to \pi K$ decays with $R_c =
2Br(\pi^0K^-)/Br(\pi^-\bar{K}^0)$ and $R_n
=Br(\pi^+K^-)/2Br(\pi^0\bar{K}^0)$ defines a sensitive quantity for
probing new physics as $\Delta R$ is dominated by the second order
of electroweak penguin contributions. A large discrepancy between
experimental data and standard model (SM) prediction $\Delta
R^{exp}/\Delta R^{SM} > 9.0\pm 5.0$  strongly indicates a signal of
new physics in the electroweak penguin sector. Within the SM, the
current $\pi K$ data favor a very large color-suppressed tree
amplitude $|C'/T'|\sim 2$, large CP violations
($A_{CP}(\pi^0\bar{K}^0)\sim 0.69$ and $A_{CP}(\pi^0\bar{K}^-)\sim
0.56$), which is connected to $\Delta R$ and be solved
simultaneously with extra electroweak penguin contributions. More
accurate measurements on the ratio difference $\Delta R$ and CP
violation in $B\to \pi \pi, \pi K$ decays may provide a window for
probing new physics and testing the isospin and SU(3) symmetries.
\end{abstract}


\maketitle

\newpage

\section{Introduction}

  The measurements of hadronic charmless $B$ decays at the two
$B$-factories become more and more accurate. Currently, all the
branching ratios of $B \rightarrow \pi \pi$ and $\pi K$ modes have
been measured with good accuracy\cite{HFAG} and a large direct CP
violation has been established in $\pi^+ K^-$ mode \cite{BB1,BB2}.
The current data show some puzzling patterns: big relative
enhancements of $\pi^0\pi^0$ and $\pi^0 \bar{K}^0$ modes, and a
large direct CP asymmetry in $\pi^+K^-$ relative to that of $\pi^0
K^-$. These are usually referred to as $\pi\pi$ and $\pi K$ puzzles.
Their implications have been investigated by many
groups\cite{BF,MY,NK,LMS,GR,CL,HM,CG,BH,IDL,LM,SB}.
In our recent paper\cite{WZ1}, it has been shown that the weak phase
$\gamma$ can well be determined to be consistent with the standard
model, a large electroweak penguin relative to tree type diagrams with
a large strong phase is preferred and an enhanced color-suppressed
tree diagram is needed. A comprehensive $\chi^2$ analysis has been
carried out including subleading diagrams, their implications to $K K$
modes as well as the SU(3) broken effects\cite{WZ2}.
In the present note, we shall present a model-independent analytical
analysis on the origins of those puzzles in the relevant decay
modes, how a signal of new physics can be singled out from the
experimental observables, and how more precise measurements can
provide a window for exploring new physics and testing isospin and
SU(3) relations from the charmless B decays.

As the experiments have already shown the possibility of large CP
asymmetries, the final state interactions (FSIs) can not be simply
neglected. It is natural to rise a question that whether the current
puzzles especially the $\pi K$ puzzle is linked to new physics
beyond the SM or just from some complicate strong interaction
effects.  The widely used quark flavor diagrams are however not
eigenstates of strong interaction which makes it difficult to get
general conclusions.  Furthermore, it is not clear whether the
usually neglected subleading diagrams of annihilation type are
indeed tiny and whether the FSIs will spoil the diagrammatic
decomposition by change the hierarchy among the diagrams
\cite{NeuGer}or introducing new diagrams like charming penguins
\cite{chmp}.  For these reasons, in the present work, we shall begin
with a general isospin analysis

The effective Hamiltonian for $\Delta S=0 (1)$ nonleptonic B
decays is given by
\beq
\label{EH} H_{eff}=\frac{G_{F}}{\sqrt{2}} \sum_{q=u,c}
\lambda_{q}^{(s)}
                       \left(
                              C_{1} O^q_{1}+ C_{2} O^q_{2} + \sum_{i=3}^{10}
                              C_{i}O_{i}
                       \right),
\eeq
 with $\lambda_{q}^{(s)} =V_{qb}V^{*}_{qd(s)}$ is the products
of CKM matrix elements. where $O^{u(c)}_{1,2}$,  $O_{3,\dots, 6}$
and $O_{7,\dots ,10}$ are related to tree, QCD penguin and
electroweak penguin sectors respectively.

The general isospin decomposition of the decay amplitudes for $B
\to \pi\pi(\pi K)$ decays can be expressed as \cite{ISO}:
\beq
 A^{\pi\pi(\pi K)}= \lambda^{(s)}_{u} A^{\pi\pi(\pi
K)}_u+\lambda^{(s)}_c A^{\pi\pi(\pi K)}_c,
 \eeq
where $\lambda^{(s)}_{u}=V_{ub}V^*_{ud(s)}$,
$\lambda^{(s)}_{c}=V_{cb}V^*_{cd(s)}$ and
\beq \label{isospinPP} A_{q}^{\pi^-\pi^+} &=&\sqrt{\frac{2}{3}}
a_0^{q}e^{i\delta_0^q} +\sqrt{\frac{1}{3}} a_2^{q}e^{i\delta_2^q},
\non
A_{q}^{\pi^0\pi^0} &=&\sqrt{\frac{1}{3}}
a_0^{q}e^{i\delta_0^q} - \sqrt{\frac{2}{3}}a_2^{q}e^{i\delta_2^q},
\non
A_{q}^{\pi^-\pi^0}&=&-\sqrt{\frac{3}{2}}a_2^{q}e^{i\delta_2^q},
\eeq
 and
 \beq
 \label{isospinPK} A_{q}^{\pi^+K^-} &=&
\sqrt{\frac{2}{3}}a_{1/2}^{q}e^{i\delta_{1/2}^q} +
\sqrt{\frac{1}{3}}a_{3/2}^{q}e^{i\delta_{3/2}^q}, \non
A_{q}^{\pi^0\bar{K^0}}&=&
\sqrt{\frac{1}{3}}a_{1/2}^{q}e^{i\delta_{1/2}^q} -
\sqrt{\frac{2}{3}} a_{3/2}^{q}e^{i\delta_{3/2}^q}, \non
A_{q}^{\pi^0K^-}&=&-\sqrt{\frac{1}{3}}b_{1/2}^{q}e^{i\delta_{1/2}^{'q}}
- \sqrt{\frac{2}{3}} a_{3/2}^{q}e^{i\delta_{3/2}^q}, \non
A_{q}^{\pi^-\bar{K^0}}&=&\sqrt{\frac{2}{3}}b_{1/2}^{q}e^{i\delta_{1/2}^{'q}}
- \sqrt{\frac{1}{3}} a_{3/2}^{q}e^{i\delta_{3/2}^q}.
 \eeq
with $a_I^q$($b_I^q$) and $\delta_I^{q}$($\delta_I^{'q}$),($q = u,c$
and $I = 0,2$ or $I = 1/2, 3/2$) the isospin amplitudes and strong
phases.

In the $B \to \pi\pi$ processes, as the effective operator with
$\Delta I = 3/2$ isospin component is unique in the effective
Hamiltonian, the hadronic matrix elements drop out in the ratio
between two isospin $I=2$ amplitudes, leading to the following relations
\cite{ISO,NR}:
    \beq
    \frac{a_2^c}{a_2^u}&\equiv &R_{EW} = \frac{2}{3}\cdot \frac{C_9+C_{10}}{C_1+C_2+C_9+C_{10}} =
(-1.25\pm 0.125)\times 10^{-2} \\
    \delta^u_{2}&= &\delta^c_{2}\equiv\delta_2
    \eeq
Note that the above relation is model independent and free from
hadronic uncertainties. A direct consequence from this relation
is  that no direct CP violation occurs in the $B\to \pi^-\pi^0$ decay, namely
 \beq
 & & A_{CP} (B\to \pi^-\pi^0) \simeq 0, \qquad \mbox{SM} \\
 & &  A_{CP} (B\to \pi^-\pi^0) \gg 0.1, \qquad \mbox{new physics}
 \eeq
as long as isospin symmetry holds at a few percent level. A similar observation was also
made within SU(3) symmetry\cite{SU3}. Obviously, the conclusion
based on isospin symmetry is  more reliable as SU(3) breaking
effects can not be neglected.
In  $B \to \pi K$ decays, the effective
operators for the highest $\Delta I = 1$ isospin components are
not unique. However, one can still define a similar ratio
\beq
R'_{EW}\equiv \frac{a^{c}_{3/2}}{a^{u}_{3/2}}.
\eeq
In SU(3) limit $R'_{EW}=R_{EW}$ and
$\delta^u_{3/2}=\delta^c_{3/2}$. Due to SU(3) breaking in the isospin
amplitudes from operators $O_1^u - O_2^u= (\bar{u} u)(\bar{s} b) -
(\bar{s} u)(\bar{u} b)$, the value of $R'_{EW}$ become slightly model
dependent. However, in the factorization approach, it has been
demonstrated that the SU(3) symmetry breaking effects are small either
because of the cancellation between two combining factors of the decay
constants and form factors, namely $f_K f_0^{B\to \pi}
\simeq f_{\pi} f_0^{B\to K}$, or  the suppression by the
heavy bottom meson mass $(m_K^2-m_{\pi}^2)/m_B^2 \ll 1$\cite{NR}.
The typical corrections are less than $10\%$. Thus, in such an
approximation, one has in the standard model
 \beq\label{REW}
   R_{EW}' \simeq R_{EW}, \qquad
    \delta_{3/2}^{u} \simeq \delta_{3/2}^{c}\equiv \delta_{3/2}
 \eeq

\section{$\pi K$ puzzle and new physics}

The relative sizes of the isospin amplitudes has been calculated in
Ref.\cite{ISO} based on factorization. From those results and
considering the most conservative uncertainties we have:
 \beq
 & & \varepsilon_{P}^{'} = \frac{a_{3/2}^{c}}{a_{1/2}^{c}}
= O(10^{-1}\sim 10^{-2}), \non
 & &  \varepsilon_{P}^{''}= \frac{a_{3/2}^{c}}{b_{1/2}^{c}}
= O(10^{-1} \sim 10^{-2}) \non
 & & R_{P}^{'}=\frac{a_{1/2}^{c}}{a_{1/2}^{u}}
=- O(10^{-1}\sim 1)\non
 & & R_{P}^{''} = \frac{b_{1/2}^c}{b_{1/2}^u}
= -O(10^{-1}\sim 1)
 \eeq
In the whole discussions bellow, we shall assume that above order
of magnitude estimate for the isospin amplitudes remain unchanged
under FSI.  With the smallness of $|\lambda_u^s|/|\lambda_c^s|
\simeq 0.02$, the branching ratios of $\pi K$ modes can be
expanded around small quantities $\varepsilon'_P$ and
$\varepsilon''_P$:
\beq R_n & = &
\frac{Br(\pi^+K^-)}{2Br(\pi^0\bar{K^0})}=\frac{1+\alpha_1^{'}\
\varepsilon_{P}^{'} + \alpha_2^{'}\
\varepsilon_{P}^{'2}}{1-\alpha_1^{'}\  \varepsilon_{P}^{'} +
4 \alpha_2^{'}\  \varepsilon_{P}^{'2}} \\ 
R_c & = & \frac{2Br(\pi^0K^-)}{Br(\pi^-\bar{K}^0)}
=\frac{1+\alpha_1^{''}\  \varepsilon_{P}^{''} + 4 \alpha_2^{'}\
\varepsilon_{P}^{''2} }{1-\alpha_1^{''}\ \varepsilon_{P}^{''}+
\alpha_2^{'}\ \varepsilon_{P}^{''2}} 
 \eeq
The coefficients have the following structure
 \beq
& & \alpha_1^{'} = \sqrt{2}
\frac{\xi_s}{R_{EW}^{'}} \left( \varepsilon_1^{'} +
\varepsilon_2^{'}/R_P^{'} \right) \non
 & & \alpha_1^{''}
= \sqrt{2} \frac{\xi_s}{R_{EW}^{'}}
\left( \varepsilon_1^{''} + \varepsilon_2^{''}/R_P^{''} \right)\non
 & & \alpha_2^{'} =  \frac{1}{2} \frac{\xi_s}{R_{EW}^{'}}
\left( \hat{\varepsilon}_1^{'} + \hat{\varepsilon}_2^{'}/R_{EW}^{'}
\right)
 \eeq
where $\xi_s=|\lambda_u^s/\lambda_c^s|$. The coefficient
$\alpha_1^{'}$($\alpha_1^{''}$) for the linear expansion is
dominated by $\varepsilon_1^{'}(\varepsilon_1^{''})$ which are
unstable under small variations of $R'_{EW}$ and $\cos\gamma$. The
expressions for $\varepsilon$ quantities are given by
 \beq
& & \varepsilon_1^{'}  = \frac{1}{\xi_s}
R_{EW}^{'} \cos{(\delta_{1/2}^{c} -\delta_{3/2}^{c})} +
\cos{\gamma} \cos{(\delta_{1/2}^{c}-\delta_{3/2}^{u})} \non
 & &
\varepsilon_1^{''}  =\frac{1}{\xi_s} R_{EW}^{'}
\cos{(\delta_{1/2}^{'c} -\delta_{3/2}^{c})} + \cos{\gamma}
\cos{(\delta_{1/2}^{'c}-\delta_{3/2}^{u})} \non
 & & \varepsilon_2^{'}
= \xi_s
\cos{(\delta_{1/2}^{u}-\delta_{3/2}^{u})}+R_{EW}^{'}\cos{\gamma}
\cos{(\delta_{1/2}^{u} - \delta_{3/2}^{c}) } \non
 & &
\varepsilon_2^{''} = \xi_s
\cos{(\delta_{1/2}^{'u}-\delta_{3/2}^{u})}+R_{EW}^{'}\cos{\gamma}
\cos{(\delta_{1/2}^{'u} - \delta_{3/2}^{c}) } \non
 & &
\hat{\varepsilon}_1^{'}  = \frac{1}{\xi_s}
R_{EW}^{'} + \cos{\gamma}
\cos{(\delta_{3/2}^{u}-\delta_{3/2}^{c})} \non
 & &
\hat{\varepsilon}_2^{'} = \xi_s
+R_{EW}^{'}\cos{\gamma} \cos{(\delta_{3/2}^{u} - \delta_{3/2}^{c})
}
 \eeq
To the next leading order of $\varepsilon'_P$ and $\varepsilon''_P$, we have
 \beq
 R_c & \simeq 1 +
2\alpha_1^{'} \varepsilon_P^{'} + (3\alpha_2^{'} - 2\alpha_1^{'2})
\varepsilon_P^{'2} \non
 R_n & \simeq  1 + 2\alpha_1^{'}
\varepsilon_P^{''} - (3\alpha_2^{'} - 2\alpha_1^{'2})
\varepsilon_P^{''2}
 \eeq
Applying  Eq.(\ref{REW}),  we  arrive at the following expressions
 \beq
& & \varepsilon_1^{''}  =  \varepsilon_1^{'}=\left(\frac{1}{\xi_s}
R_{EW}^{'}  + \cos{\gamma} \right) \cos{(\delta_{1/2}^{c}
-\delta_{3/2})} \non
 & & \varepsilon_2^{'} = \left(\xi_s +R_{EW}^{'}\cos{\gamma} \right)
\cos{(\delta_{1/2}^{u}-\delta_{3/2})} \non
 & & \varepsilon_2^{''} = \left( \xi_s +R_{EW}^{'}\cos{\gamma} \right)
\cos{(\delta_{1/2}^{u}-\delta_{3/2})}
 \non
 & & \hat{\varepsilon}_1^{'}  = \frac{1}{\xi_s}
R_{EW}^{'} + \cos{\gamma} \non 
&  & \hat{\varepsilon}_2^{'} = \xi_s +R_{EW}^{'}\cos{\gamma}
 \eeq
Taking $\xi_s\simeq 0.02$ and $\cos\gamma\simeq 60^\circ$
\beq
\frac{1}{\xi_s} R_{EW}^{'} + \cos{\gamma}\simeq -0.125, \quad
\xi_s
+R_{EW}^{'}\cos{\gamma} \simeq 0.014
\eeq
Note that a model-independent cancelation takes place in the first
factor which greatly suppresses the coefficients $\alpha'_1$ and
$\alpha''_1$.  As a consequence, large deviations of $R_n$ and $R_c$
at $\sim 10\%$ from unity becomes unlikely. For in SM, with the most
conservative estimation, we can get:
\beq & & \alpha_1^{'} \simeq 0.28 \cos{(\delta_{1/2}^{c}
-\delta_{3/2})}, \quad \non & & \alpha_2^{'} \simeq 1.0 \non
 & & \varepsilon_{P}^{'}\simeq \varepsilon_{P}^{''}\simeq 0.04\sim 0.06
\eeq
As a consequence, we obtain the following
correlated numerical predictions for the ratios
\beq
0.97<R_c<1.05, \qquad 0.96<R_n<1.04
\eeq
and
\beq \Delta R^{SM} \simeq 2(3\alpha_2^{'} - 2\alpha_1^{'2})
\varepsilon_P^{'2} \simeq 0.01\sim 0.02
 \eeq
which do not agree well with the current data
\beq R_c^{ex}=1.0\pm0.06, \quad R_n^{ex} = 0.82\pm0.08, \quad
\mbox{or} \ \ \Delta R^{ex} =0.18 \pm 0.10
\eeq
The ratio $R_n^{exp}$ or the ratio difference $\Delta R$ is about
$2\sigma$ away from the standard model predictions. This may be
regarded as an indication for new physics beyond the standard model.

The above observations are obtained from flavor symmetries and
depends mostly on isospin. It is expected that it is more robuster
comparing with diagrammatic method which based on short distance
pictures. To see the implications in terms of quark flavor flow
diagrams, one may rewrite the isospin amplitudes as follows:

\beq
  a_0^{u}e^{i\delta_0^{u}} &=&
  -\frac{1}{\sqrt{6}}[\hat{T} - 3\hat{P} + (\hat{T}-3\bar{C}) + (\hat{P}_{EW}-3 P_{EW}^C)],\non
  a_0^{c}e^{i\delta_0^{c}} &=&
  \frac{1}{\sqrt{6}}[3\hat{P} - (\hat{P}_{EW}-3 P_{EW}^C) ],\non
  a_2^{u}e^{i\delta_2^{u}} &=&
  -\frac{1}{\sqrt{3}}[\hat{T}-\hat{P}_{EW}],\non
  a_2^{c}e^{i\delta_2^{c}} &=&
  \frac{1}{\sqrt{3}}\hat{P}_{EW}
  \eeq
and similarly
  \beq
  a_{1/2}^{u}e^{i\delta_{1/2}^{u}} &=&
  -\frac{1}{\sqrt{6}}[\hat{T}^{'} -3\hat{P}^{'} + (\hat{T}^{'}-3C^{'}) + (\hat{P}_{EW}^{'}-3 P_{EW}^{C'}) ],\non
  a_{1/2}^{c}e^{i\delta_{1/2}^{c}} &=&
  \frac{1}{\sqrt{6}}[ 3\hat{P}^{'}-(\hat{P}_{EW}^{'}-3 P_{EW}^{C'})  ],\non
  a_{3/2}^{u}e^{i\delta_{3/2}^{u}} &=&
  -\frac{1}{\sqrt{3}}[\hat{T}^{'}-\hat{P}_{EW}^{'}],\non
  a_{3/2}^{c}e^{i\delta_{3/2}^{c}} &=&
  \frac{1}{\sqrt{3}}\hat{P}_{EW}^{'},\non
  b_{1/2}^ue^{i\delta_{1/2}^{'u}} &=&
  -\frac{1}{\sqrt{6}}[\hat{T}^{'} -(3\hat{P}^{'}+ \hat{P}'_{EW} ) -3A' ],\non
  b_{1/2}^ce^{i\delta_{1/2}^{'c}} &=&
  \frac{1}{\sqrt{6}}(3\hat{P}^{'}+ \hat{P}'_{EW} )
  \eeq
 Where $\hat{T} = T + C = \bar{T} + \bar{C}$ ,
$\bar{T} = T + E$, $\bar{C} = C - E$, $\hat{P} = P -P_{EW}^C/3 +
P_A$, $\hat{P}_{EW} = P_{EW} + P_{EW}^C$, and similarly $\hat{T}'
= T' + C'$, $\hat{P}^{'} = P^{'} -P_{EW}^{C'}/3$, $\hat{P}_{EW}'=
P_{EW}' + P_{EW}^{C'}$. Here $T$,$C$,$P$,$P_{EW}$,$P_{EW}^C$
represent the tree diagram, color-suppressed tree diagram, QCD
penguin diagram, electroweak penguin diagram and color suppressed
electroweak penguin diagram, and $E$,$A$ and $P_A$ denote the
exchange diagram, annihilation diagram and penguin annihilation
diagram respectively.

In the diagrammatic language, the counterpart of Eq.(\ref{REW}) is
given by
 \beq
  R^{(')}_{EW}\simeq
\frac{\hat{P}^{(')}_{EW}}{\hat{T}^{(')}}
 \eeq
  and
 \beq
 R_n &\simeq&
 1+(r_{EWC})^2+(r_{EW})^2(4\cos{\delta_{EW}^{'2}}-1)-2(r_{EWC}\cos{\delta_{EW}^{C'}}+r_{EW}\cos{\delta_{EW}'})\non
 &&-2\xi_sr_T\cos{\gamma}\cos{\delta_{T}'}+2\xi_sr_Tr_{EWC}\cos{\gamma}\cos{(\delta_{T}'-\delta_{EW}^{C'})}\non
 &&-2\xi_sr_C\cos{\gamma}\cos{\delta_{C}'}-2\xi_sr_Cr_{EW}\left[\cos{\gamma}\cos{(\delta_{C}'-\delta_{EW}')}
 -4\cos{\gamma}\cos{\delta_C'}\cos{\delta_{EW}'}\right]\non
 &&+4\xi_sr_Tr_{EW}\cos{\gamma}\cos{\delta_T'}\cos{\delta_{EW}'}+4\xi_sr_Cr_{EWC}\cos{\gamma}\cos{\delta_C'}\cos{\delta_{EW}^{C'}}\non
 &&+4r_{EWC}r_{EW}\cos{\delta_{EW}^{C'}}\cos{\delta_{EW}'},\non
 R_c &\simeq&
 1+(r_{EWC})^2+(r_{EW})^2-2(r_{EWC}\cos{\delta_{EW}^{C'}}+r_{EW}\cos{\delta_{EW}'})\non
 &&-2\xi_sr_T\cos{\gamma}\cos{\delta_{T}'}+2\xi_sr_Tr_{EWC}\cos{\gamma}\cos{(\delta_{T}'-\delta_{EW}^{C'})}\non
 &&-2\xi_sr_C\cos{\gamma}\cos{\delta_{C}'}+2\xi_sr_Cr_{EW}\cos{\gamma}\cos{(\delta_{C}'-\delta_{EW}')}\non
 &&+2\xi_s\cos{\gamma}\left[r_{EWC}r_C\cos{(\delta_{EW}^{C'}-\delta_C')}+r_{EW}r_T\cos{(\delta_{EW}'-\delta_T')}\right]\non
 &&+2r_{EWC}r_{EW}\cos{(\delta_{EW}'-\delta_{EW}^{C'})},
\eeq
To get the above expressions, we have neglected the terms
proportional to $\xi_s^2$. Here $r_F = |F'/\hat{P}'|$ with F denote
Feynman diagrams ($F'=T',C', P_{EW}'$ etc). We define $F' =
|F'|e^{i\delta_F}$ except for the E-W penguin $P_{EW}^{(C)'} =
-|P_{EW}^{(C)'}|e^{i\delta_{EW(C)}'}$ in the rest of this paper. The
difference between $R_n$ and $R_c$ is of the second order
 \beq
 \Delta R&\simeq&
 -2(r_{EW})^2\cos{(2\delta_{EW}')}-2r_{EW}r_{EWC}\cos{(\delta_{EW}'+\delta_{EW}^{C'})}\non
 &&-2\xi_sr_Cr_{EWC}\cos{\gamma}\cos{(\delta_{C}'+\delta_{EW}^{C'})}\non
 &&-2\xi_sr_Tr_{EW}\cos{\gamma}\cos{(\delta_{T}'+\delta_{EW}')}\non
 &&-4\xi_sr_Cr_{EW}\cos{\gamma}\cos{(\delta_{C}'+\delta_{EW}')}
  \eeq
 The third to fifth terms are all $\xi_s$ suppressed and $\Delta R$ is
dominated by electroweak penguin and very sensitive to any new
physics in electroweak penguin sector. It is not difficult to
check that in order to simultaneously explain the experimental
data, it requires that
 \beq
 \alpha_1^{'} \varepsilon_P^{'} \simeq - 0.05, \qquad \alpha_2^{'} \simeq
 14 \alpha_1^{'2}
 \eeq
which leads to the following solution
 \beq
  & &  \delta_{3/2}^{u}-\delta_{3/2}^{c} \simeq - \pi/3 ,\qquad \delta_{1/2}^{c} -\delta_{3/2}^{u} \simeq 0.40
  \non
  & & \alpha_2^{'} \simeq 0.68, \qquad \alpha_1^{'} \simeq -0.22
  \non
 & & \varepsilon_P^{'} \simeq 0.20, \quad \mbox{i.e.} \quad  |\hat{P}_{EW}'/\hat{P}'| \simeq
 0.42
 \eeq
 For the typical values of $|\hat{P}'/\hat{T}'| = 0.10\sim 0.15$, we
 find
 \beq
& &  R_{EW}^{'}|_{exp}
= -(0.04\sim 0.06) \\
& & (\delta_{3/2}^{u}-\delta_{3/2}^{c})|_{exp} \simeq - \pi/3
 \eeq
which shows that a large enhancement by a factor 3.5 to 5 for the
effective electroweak penguin contributions is needed and also a
large strong phase difference is required. This may imply the
existence of new physics with a significant contribution to the
effective electroweak penguin operator.

With the above analytical analysis, we can arrive at the following
conclusions:
 \beq
 & & \Delta R = R_c - R_n \leq 0.02, \qquad \mbox{ isospin $\&$ SU(3)
 relations in SM }\\
 & & \Delta R = R_c - R_n > 0.02, \qquad \mbox{ SU(3) symmetry breaking/new
 physics} \\
 & & \Delta R = R_c - R_n \gg 0.02,  \qquad \mbox{ signal of new
 physics}
 \eeq
The present experimental data for $\Delta R$ is much larger than
the standard model prediction with using isospin and SU(3)
relations, i.e.,
 \beq
 (\Delta R)^{exp}/(\Delta R)^{SM} > 9.0 \pm 5.0
 \eeq
which indicates a signal of new physics. It is clear that more
precise measurements on the difference between two ratios will
provide an effective way to probe new physics beyond the standard
model.

An alternative possibility is to introduce a new CP phase
$\phi_{NP}$ in EW penguin diagram, which was discussed recently in
ref.\cite{BFRS}. Taking the isospin relation for the amplitude
$R_{EW}' = R_{EW} = -0.0125$, but with introducing a new weak phase
$\phi_{NP}$ for the EW penguin. As a consequence, the expressions in
Eq(15) will be modified as:
 \beq
& & \varepsilon_1^{'}  = \frac{1}{\xi_s} R_{EW}^{'}\cos{\phi_{NP}}
\cos{(\delta_{1/2}^{c} -\delta_{3/2}^{c})} + \cos{\gamma}
\cos{(\delta_{1/2}^{c}-\delta_{3/2}^{u})} \non
 & &
\varepsilon_1^{''}  = \frac{1}{\xi_s} R_{EW}^{'}\cos{\phi_{NP}}
\cos{(\delta_{1/2}^{'c} -\delta_{3/2}^{c})} + \cos{\gamma}
\cos{(\delta_{1/2}^{'c}-\delta_{3/2}^{u})} \non
 & & \varepsilon_2^{'}
= \xi_s
\cos{(\delta_{1/2}^{u}-\delta_{3/2}^{u})}+R_{EW}^{'}\cos{(\gamma+\phi_{NP})}
\cos{(\delta_{1/2}^{u} - \delta_{3/2}^{c}) } \non
 & &
\varepsilon_2^{''} = \xi_s
\cos{(\delta_{1/2}^{'u}-\delta_{3/2}^{u})}+R_{EW}^{'}\cos{(\gamma+\phi_{NP})}
\cos{(\delta_{1/2}^{'u} - \delta_{3/2}^{c}) } \non
 & &
\hat{\varepsilon}_1^{'}  = \frac{1}{\xi_s} R_{EW}^{'} +
\cos{(\gamma+\phi_{NP})} \cos{(\delta_{3/2}^{u}-\delta_{3/2}^{c})}
\non
 & &
\hat{\varepsilon}_2^{'} = \xi_s
+R_{EW}^{'}\cos{(\gamma+\phi_{NP})} \cos{(\delta_{3/2}^{u} -
\delta_{3/2}^{c}) }
 \eeq
As $\phi_{NP}$ is a free parameter, thus
$\cos{(\gamma+\phi_{NP})}$ can be $[-1,1]$.  We then obtain the
following solution in order to explain the experimental data
within $1\sigma$ error:
 \beq
  & &  \delta_{3/2}^{u} = \delta_{3/2}^{c}, \quad \delta_{1/2}^{c}-\delta_{3/2}^{c} \simeq \pm1.3 ,\non
  & & \alpha_2^{'} \leq 3.4, \qquad \alpha_1^{'} \leq -0.5,
  \non
 & & \varepsilon_P^{'} \geq 0.06,
 \eeq
where the smallest $\varepsilon_P^{'}$ occurs at
$\cos{(\gamma+\phi_{NP})} = -1$. It shows that a new CP phase can
truly bring the results closer to the present data without a large
enhanced E-W penguin .

From the above discussions,  the $R_n$ and $R_c$ puzzle can be
solved through new physics in the EW penguin sector with an enhanced
amplitude and/or a new CP-violating phase. In general, both effects
can improve the discrepancy between the present data and the SM
predictions.

\section{Enhanced Color-Suppressed Amplitudes from $B\to \pi\pi$}

We now consider the $B\to \pi\pi$ decays  using the diagrammatic
method, the CP-averaging branching ratios have the following
forms:
 \beq
\frac{1}{\tau_{B^0}}Br(\pi^+\pi^-) &=&
|\lambda_u|^2|\bar{T}|^2+(|\lambda_u|^2+|\lambda_c|^2-2\cos{\gamma}|\lambda_u||\lambda_c|)|\bar{P}|^2\non
   &&+2|\lambda_u||\bar{P}||\bar{T}|\cos{\delta_{\bar{T}}}(|\lambda_c|\cos{\gamma}-|\lambda_u|),\non
 \frac{1}{\tau_{B^0}}Br(\pi^0\pi^0) &=& \frac{1}{2}
 [|\lambda_u|^2|\bar{C}|^2+(|\lambda_u|^2
 +|\lambda_c|^2-2\cos{\gamma}|\lambda_u||\lambda_c|)|\bar{P}-\hat{P}_{EW}|^2]\non
 &&-2|\lambda_u||\bar{P}-\hat{P}_{EW}||\bar{C}|\cos{\delta_{\bar{C}}}(|\lambda_c|\cos{\gamma}-|\lambda_u|),\non
\frac{1}{\tau_{B^-}}Br(\pi^-\pi^0) &=&
\frac{1}{2}|\lambda_u|^2|\bar{T}+\bar{C}|^2
+(|\lambda_u|^2+|\lambda_c|^2-2\cos{\gamma}|\lambda_u||\lambda_c|)|\hat{P}_{EW}|^2\non
   &&-2|\lambda_u||\hat{P}_{EW}||\bar{T}+\bar{C}|\cos{(\delta_{\bar{T}+\bar{C}}-\delta_{EW})}(|\lambda_c|\cos{\gamma}-|\lambda_u|)
\eeq
where $\bar{P} = \hat{P} + P_{EW}^c$ and
$\delta_{\bar{C}},\delta_{\bar{T}},\delta_{EW},\delta_{\bar{T}+\bar{C}}$
are the strong phases of $\bar{C}$, $\bar{T}$, $\hat{P}_{EW}$ and
$\bar{T}+\bar{C}$. We fix the strong phase of
$\hat{P}(\delta_{\hat{P}})$ to be zero as an overall phase. Noticing
the fact that $|\bar{P}| \ll |\bar{T}|$ and
$2|\lambda_u^d|(|\lambda_c^d| \cos{\gamma}-|\lambda_u^d|) \approx
0.4 |\lambda_u^d|^2$, we obtain in a good approximation the
following relations
 \beq
\frac{R_-}{(1-R_0)} \approx \frac{1 +|\bar{C}/\bar{T}|^2 + 2
 |\bar{C}/\bar{T}|\cos{(\delta_{\bar{T}}-\delta_{\bar{C}})}}{1
 -|\bar{C}/\bar{T}|^2}
\eeq with $R_{-}\equiv 2Br(\pi^-\pi^0)/\tau Br(\pi^+\pi^-)$ and $
R_0 \equiv 2Br(\pi^0\pi^0)/Br(\pi^+\pi^-)$. where $\tau=
\tau_{B^-}/\tau_{B^0} = 1.086$ reflects the life-time difference.
Taking the experimental data for the three branching ratios and
considering the possible range for
$\cos{(\delta_{\bar{T}}-\delta_{\bar{C}})}\in [1,-1]$, we arrive at
the following constraint for the ratio $|\bar{C}|/|\bar{T}|$ \beq
 0.68 \leq \frac{|\bar{C}|}{|\bar{T}|}\leq 0.98
 \eeq
 Noticing the positivity of the quantity
 \[
(|\lambda_u|^2+|\lambda_c|^2-2\cos{\gamma}|\lambda_u||\lambda_c|)|\bar{P}|^2
+2|\lambda_u||\bar{P}||\bar{T}|\cos{\delta_{\bar{T}}}(|\lambda_c|\cos{\gamma}-|\lambda_u|)
> 0 \]
namely, $ Br(\pi^+\pi^-)/\tau_B > |\lambda_u|^2 |\bar{T}|^2$, we
yield a more strong constraint for the ratio
 \beq
\frac{|\bar{C}|}{|\bar{T}|}\leq \sqrt {R_{0}} \equiv
\sqrt{\frac{2Br(\pi^0\pi^0)}{Br(\pi^+\pi^-)} }\simeq 0.76
 \eeq
Combining the above two constraints,
 \beq
 0.68 \leq |\bar{C}|/|\bar{T}| \leq 0.76 ,\quad \mbox{or} \quad
|\bar{C}|/|\bar{T}|= 0.72\pm 0.04,
 \eeq
we obtain from eq.(38) the following allowed ranges for two
amplitudes $\bar{T}$ and $\bar{C}$ and the difference of their
strong phases
 \beq
 & & |\bar{T}| \simeq 0.58 \pm 0.02, \non
 & & |\bar{C}| \simeq 0.41\pm 0.03,\non
 & & \cos{(\delta_{\bar{T}}-\delta_{\bar{C}})} \simeq 0.70 \mp 0.30
 \eeq
Note that the above numerical values are obtained for simplicity by
only taking the central values of the experimental data. When taking
into account the experimental errors, the allowed range could be
enlarged by $(10 \sim 20)\%$. While the above results are almost not
affected by the enhanced electro-weak penguin contributions as long
as $|\hat{P}_{EW}| \ll |\bar{T}|$ remains a good approximation.
Obviously, the resulting ratio $|\bar{C}|/|\bar{T}| \simeq 0.72 $ is
much larger than the result$ |\bar{C}|/|\bar{T}| \simeq 0.1\sim 0.2
$ calculated from both the QCD factorization approach\cite{QCDF} and
perturbative QCD approach\cite{PQCD}. Although the recent next to
leading order QCD factorization calculations show some enhancement
of $C$, it is still difficult to meet the current
data\cite{QCDFnlo}, and a large color suppressed tree diagram is
also independently favored by $\pi K$ and $K \eta^{(')}$ data
\cite{BH,WZ2,Keta}.

\section{Large $|C'/T'|$ and Large Electroweak Penguin }

Let us now discuss the so-called large $|C'/T'|$ puzzle in the
$B\to \pi K$ decays. Firstly, considering the case with the SU(3)
relations $P'_{EW} \simeq R'_{EW} T'$, $P_{EW}^{C'} = R'_{EW} C'$
and $R'_{EW} \simeq R_{EW} \simeq -0.0125$(refer to Case I). With
this consideration, the numbers of parameters in the $B\to \pi K$
decays are reduced to be five, i.e., $|T'|$, $|C'|$, $|\hat{P}'|$,
$\delta_{T'}$ and $\delta_{C'}$ (The strong phase of $\hat{P}'
(\delta_{P'} = 0)$ as an overall phase). They can be solved by
five experimental data, namely four branching ratios and one
direct CP violation of $B\to \pi^+ K^-$ decays. For simplicity,
taking the central values of experimental data, we found in this
case it's impossible to find a solution to meet all the five data
points. If we only use four data points of branching ratios in
$\pi K$ system in Case I analysis, numerical results are then
found with SU(3) relations and $\gamma \simeq \pi/3$ to be
 \beq
 & & |T'| \simeq 1.20, \qquad \delta_{T'} \simeq 0.30 \non
 & & |C'/T'| \simeq (2.0\sim 3.0), \qquad \delta_{C'} \simeq 2.2 \non
 & & |\hat{P}'| \simeq 0.12,   \qquad  \mbox{(Case I)}
 \eeq
The resulting ratio $|C'/T'|$ is unexpected large compared with QCD
factorization and perturbative QCD approaches. It is also larger by
a factor of three than that extracted from $\pi\pi$ modes
$|\bar{C}/\bar{T}|\simeq 0.72$. This confirms the previous
observations in Ref.\cite{BH,WZ2}.

It is natural to ask whether the above puzzle is related to the
ratio difference $\Delta R = R_c-R_n$, and can simultaneously be
solved by new contributions from electroweak penguin. To check
that, first taking an enhanced ratio $R'_{EW} \simeq -0.04$, but
with the strong phase $\delta_{P'_{EW}}$ as a free parameter and
also neglecting $P_{EW}^{C'}$ (refer to Case II), we find the
following results which can meet all the five data points in
$1\sigma$ error:
 \beq
 & & |T'| \simeq 1.07, \quad \delta_{T'} \simeq 0.31, \quad |\hat{P}'| \simeq 0.12, \non
 & & |C'/T'| \simeq 0.60\sim 0.80, \quad  \delta_{C'} \simeq \pm (2.0\sim 2.5), \non
 & &\quad \mbox{for} \quad \delta_{EW}' \simeq \pm (1.4\sim 1.5), \quad  \mbox{(Case II)}
 \eeq
The ratio $|C'/T'|$ is found to be sensitive to the strong phase
$\delta_{EW}'$ of electroweak penguin. It is seen that a large
value of electroweak penguin can reduce the color-suppressed tree
amplitude $C'$. While the resulting value for the tree amplitude
$T'$ remains somehow larger than $\bar{T}$ extracted from $B\to
\pi\pi$ decays, which may indicate that the contribution from the
color-suppressed electroweak penguin $P_{EW}^{C'}$ may not be
neglected and it could be enhanced in a similar way. To see that,
further taking the following relation (refer to Case III)
 \beq
| \frac{P_{EW}^{C'}}{P'_{EW}}| \simeq |\frac{C'}{T'}| \simeq 0.70
 \eeq
it is then not difficult to find that when appropriately taking
the strong phases, the tree amplitude $T'$ can truly be largely
reduced to a low value with the following typical solution
 \beq
 & & |T'| \simeq 0.60, \qquad  \delta_{T'} \simeq 0.5 , \qquad |C'/T'| \simeq 0.60\sim 0.80, \qquad |P'|
 \simeq 0.12 \non
 & & \delta_{C'} \simeq \pm (2.0 \sim 2.4), \quad
 \delta_{EW}'\simeq \pm(1.4 \sim 1.6), \quad  \delta_{EW}^{C'} \simeq 1.3 \sim 1.5,  \quad  \mbox{(Case III)}
 \eeq

It is noticed that the solution is sensitive to the strong phases
$\delta_{EW}'$ and $\delta_{EW}^{C'}$. The reason is simple that
the tree amplitude $T'$ and the color-suppressed electroweak
penguin amplitude $P_{EW}^{C'}$ are associated with the small and
large CKM factors $\lambda_u^{s}$ and $\lambda_c^{s}$
respectively, here $|\lambda_u^{s}/ \lambda_c^{s}| \simeq 0.02$.
Thus as long as $|P_{EW}^{C'}/T'| > 0.02$ and $|P'_{EW} / C'| >
0.02$, the contributions from the amplitudes $P_{EW}^{C'}$ and
$P'_{EW}$ become sizable and more significant than the ones from
the amplitudes $T'$ and $C'$ respectively.

Similar to the analysis of $R_n$ and $R_c$, we consider the fourth
case with an additional weak phase $\phi_{NP}$ in the electroweak
penguin, but keeping the amplitude obtained from the isospin
relation $P_{EW}' = R_{EW}T', P_{EW}^{C'} = R_{EW}C'$ (refer to
Case IV). If we take a typical value $\phi_{NP} = \pi/2$ and only
use four data points of branching ratios, we get the following
solution:
 \beq
& & |T'| \simeq 1.4, \quad |C'/T'| \simeq 1.0,\qquad \delta_{T'} =
\delta_{EW}' \simeq 1.0 , \qquad |P'|
 \simeq 0.12 \non
 & & \delta_{C'} =  \delta_{EW}^{C'} \simeq 2.0, \quad \quad  \mbox{(Case IV)}
 \eeq

It is found that this new weak phase $\phi_{NP}$ in E-W penguin
sector can't reduce much larger $|T'|\simeq 1.4$ and larger ratio
$|C'/T'| \simeq 1.0$ than $|\bar{T}|\simeq 0.58$ and
$|\bar{C}/\bar{T}| \simeq 0.7$ we got in $\pi\pi$ system if SU(3)
symmetry is not broken strongly. So it is not enough to solve the
large $|C'/T'|$ puzzle by merely introducing the weak phase though
it helps to solve the large $R_n-R_c$ puzzle.

It has recently been shown\cite{LMS} by using perturbative QCD
approach that the next-to-leading-order contributions from the
vertex corrections of the quark loops and the magnetic penguins
may also provide a solution for the puzzle of very large
$|C'/T'|$. While the resulting ratio difference $\Delta R$ remains
much smaller than the experimental data, which is actually
expected from our above analyzes for a conclusion that all the
standard model predictions lead to a small $\Delta R$. This is
attributed to the smaller PQCD results for the $B^0\to \pi^0 K^0$
branching ratio and hence for a large ratio $R_n$.

\section{Implications of CP Violation}

To be consistent with the experimental results and keeping the
isospin and SU(3) relations for the electroweak penguin amplitudes
and phases, i.e., Case I, besides the
puzzle for the very large ratio $|C'/T'|\sim 2 $ and the
discrepancy with the experimental data for the branching ratio
difference $\Delta R^{exp}$,  the resulting CP asymmetries in $B\to \pi^0
\bar{K}^0$ and $B^-\to \pi^0 \bar{K}^-$ are also much larger than
the experimental data. The CP asymmetry can be expressed as
follows
 \beq
 \frac{1}{\tau_{B^0}} A_{CP}(B \to \pi^+ K^-) & \cdot & Br(B \to \pi^+ K^-) \non
  & = & -2|\lambda_u^s\lambda_c^s|\sin{\gamma} |T'| \left[|\hat{P'}|\sin{\delta_{T'}}
  -|P_{EW}^{C'}|\sin{(\delta_{T'}-\delta_{EW}^{C'})}\right],\non
\frac{1}{\tau_{B^0}}A_{CP}(B \to \pi^0 \bar{K^0}) & \cdot & Br(B \to
\pi^0 \bar{K^0}) \non
 &= & |\lambda_u^s\lambda_c^s|\sin{\gamma}|C'|
[|\hat{P'}|\sin{\delta_{C'}}+|P_{EW}'|\sin{(\delta_{C'}-\delta_{EW'})}],\\
\frac{1}{\tau_{B^-}}A_{CP}(B \to \pi^0 K^-)& \cdot & Br(B \to \pi^0
K^-) = -|\lambda_u^s\lambda_c^s|\sin{\gamma} \non
 & \cdot & [ |T'|(|\hat{P'}|\sin{\delta_{T'}}
                        -|P_{EW}'|\sin{(\delta_{T'}-\delta_{EW'})}
                        - |P_{EW}^{C'}|\sin{(\delta_{T'}-\delta_{EW}^{C'})}) \non
                        && + |C'| (|\hat{P'}|\sin{\delta_{C'}} -|P_{EW}'|\sin{(\delta_{C'}-\delta_{EW'})}
                        - |P_{EW}^{C'}|\sin{(\delta_{C'}-\delta_{EW}^{C'})})] \nonumber
 \eeq
 and the time-dependent CP asymmetry for $B^0 \to K_S\pi^0$ is
 defined as:
 \beq
{\cal A}_{K_S\pi^0}(t) &\equiv& \frac{\Gamma(\bar{B}^0(t) \to
K_S\pi^0)-\Gamma(B^0(t) \to K_S\pi^0)}{\Gamma(\bar{B}^0(t) \to
K_S\pi^0)+\Gamma(B^0(t) \to K_S\pi^0)}\non
 && \equiv S_{K_S\pi^0}\sin{(\Delta m_B t)} - C_{K_S\pi^0}\sin{(\Delta m_B t)},
 \eeq
 The $S_{K_S\pi^0}$ and $C_{K_S\pi^0}=-A_{CP}(K_S\pi^0)$ are mixing-induced and direct CP-violating
 parameters respectively. The expression for $S_{K_S\pi^0}$ is
 given by:
 \beq
 S_{K_S\pi^0} &\simeq& \sin{(2\beta)} +
 2  r_C'  \cos{(2\beta)}\cos{\delta_{C'}}\sin{\gamma}-
 2  r_C'^2 \sin{(2\beta)}\sin^2{\gamma}\non
 &&- r_C'^2 \cos{(2\beta)}\cos{(2\delta_{C'})}\sin{(2\gamma)}  - 2
  r_C' r_{EW} \cos{(2\beta)}\cos{(\delta_{C}'+\delta_{EW}')}\sin{\gamma},
 \eeq
Here $r_C'\simeq \xi_s|C'/\hat{P'}|$, and $r_{EW} =
|P_{EW}'/\hat{P'}|$. In our numerical calculations, we will use
the latest experimental result for $\sin{(2\beta)}=0.687 \pm
0.032$ as an input parameter. It is known that the experimental
result for $S_{K_S\pi^0}$ has a significant deviation from
$\sin{(2\beta)}$, it should be attributed to the subleading terms
relevant to $r_C'$ and $r_{EW}$. Numerically, it is found that in
Case I the CP asymmetries are given by
 \beq
 A_{CP}(\pi^0\bar{K^0}) \sim 0.69,\qquad
  A_{CP}(\pi^0\bar{K^-}) \sim  0.56
\eeq
 which are too large in comparison with the latest experimental
results\cite{HFAG} $A_{CP}(\pi^0\bar{K^0})= 0.02\pm 0.13$(BABAR
and Belle's results have opposite
sign:$-0.06\pm0.18\pm0.03$(BABAR);$0.11\pm0.18\pm0.08$(Belle) )
and $A_{CP}(\pi^0\bar{K}^-) = 0.04 \pm 0.04 $ .

It is then natural to ask whether the enhanced electroweak penguin
amplitude can simultaneously solve the above puzzle. Firstly,
consider the Case II, namely $R'_{EW} \simeq -0.04$ but with the
strong phase $\delta_{EW}'$ as a free parameter and also
neglecting the color-suppressed electroweak penguin contribution
$P^{C'}_{EW}$. The resulting CP violation in this case has two
possible solutions for $\delta_{EW}' > 0$
 \beq
 & & A_{CP}(\pi^0\bar{K}^0)\simeq -0.12 , \quad \mbox{$\delta_{C'} < 0$};
 \quad -0.25 , \quad \mbox{$\delta_{C'} > 0$} \\
 & & A_{CP}(\pi^0\bar{K}^-) \simeq 0.02 , \quad \mbox{$\delta_{C'} < 0$};
 \quad 0.12 , \quad \mbox{$\delta_{C'} > 0$}
\eeq
 the $\delta_C' < 0$ solution is consistent with the experimental data at $1\sigma$
 level. The other two solutions for $\delta_{EW}' < 0$ are not
 consistent with the data.

We now consider the Case III, i.e., including $P^{C'}_{EW}$ with
$|P_{EW}^{C'}/P'_{EW}| \simeq |C'/T'| \simeq 0.70$. In this case,
it is interesting to notice that only one solution with
$\delta_{EW}' > 0$ and $\delta_{C'} < 0$ is consistent with the
experimental data and the corresponding CP asymmetry is:
 \beq
 A_{CP}(\pi^0\bar{K}^0) \simeq -0.08 ,\qquad
  A_{CP}(\pi^0\bar{K}^-) \simeq -0.02
\eeq

It is seen that including the constraint of CP violation not only
helps to reduce the ambiguities for the signs of strong phases,
but also provides a consistent check for a signal of new physics.
All of the puzzles likely indicate new physics in the electroweak
penguin sector.

To check the SU(3) relations with symmetry breaking effects only
for the amplitudes $T,P,C$, we further evaluate the CP violation
in $B\to \pi\pi$ decays.  Neglecting the contributions from the
annihilation amplitudes $E$ and $P_A$ which are found to be small
from the analysis of $B \to KK$ decays, we then have in the Case
III that:
 \beq
A_{CP}(B \to \pi^+\pi^-) &\simeq& 0.41  , \non
 A_{CP}(B \to \pi^0\pi^0) &\simeq& 0.57,
 \eeq
which are consistent with experimental data at $1\sigma$
level\cite{HFAG}: $A_{CP}^{exp}(B \to \pi^+\pi^-) = 0.37\pm 0.10
 $ and $A_{CP}^{exp}(B \to \pi^0\pi^0) = 0.28\pm 0.40$.

Considering Case IV with a new CP phase ($\phi_{NP} = \pi/2$) of
EW penguin without changing the isospin relation. For in this
case, $|C'/T'|$ are still large from analysis of branching ratios
of $B \to \pi K$,which leads to a bigger $|P_{EW}^{C'}|$ and there
will be new terms in the expressions of $A_{CP}$ such as
$|\hat{P}'P_{EW}'|\sin{\phi_{NP}}\sin{\delta_{EW}'}$ term, our
calculation shows that in this case the direct CP violations are
not consistent with the data.

\begin{table}[t]
\begin{center}
\caption{Mixing induced CP violation (The four columns refer to I-IV
cases mentioned above)} \vspace{0.2cm}
\begin{tabular} {c|c|c|c|c|c}  \hline
Case &   I &  II
& III &   IV & Exp.\cite{HFAG}  \\
\hline 
$S_{\pi^0 K_S}$ &$0.32 $&$0.62$&$0.68$&$0.50$&$0.31\pm0.26$\\
\hline   
$S_{\pi^0\pi^0}$ &$-$ &$-$&$-0.70$&$-$&$-$\\
\hline   
$S_{\pi^+\pi^-}$ &$-$&$-$&$-0.63$&$-$&$-0.50\pm0.12$\\
\hline   
\end{tabular}\end{center}
\end{table}

Finally, we also check the mixing induced CP asymmetries in $B \to
\pi^0\pi^0, \pi^+\pi^-, \pi^0 K_S$ decays. The results are shown in
Table I. It is seen that in Case I(the first column), $S_{K_S\pi^0}$
coincides with the measured value as in this case $|C'/P'|$ is large
enough to give a comparable cancelation to $\sin{(2\beta)}$. And in
Case IV, a new CP phase $\phi_{NP} = \pi/2$ can also take the
$S_{K_S\pi}$ closer to the experimental data. In other cases with
new electroweak penguin contributions, larger mixing-induced CP
violation occur, but it is still consistent with the data at
$1.5\sigma$ level. For only the results of $|C'|, |T'|$ in Case III
that extracted from the $\pi K$ decays are consistent with the
results in $\pi\pi$ system within the $SU(3)$ symmetry or even
considering a little breaking effects, we just calculate the
mixing-induce CPV of $\pi\pi$ system in Case III .

\section{Conclusions}

In summary, we have presented a model-independent analytical
analysis for charmless B decays. The quantity $\Delta R = R_c -
R_n$ defined by four CP-averaging branching ratios in $B\to \pi K$
decays with $R_c = 2Br(\pi^0K^-)/Br(\pi^-\hat{K}^0)$ and $R_n
=Br(\pi^+K^-)/2Br(\pi^0\hat{K^0})$ provides a promise way for
probing new physics beyond the standard model. This is because the
ratio difference $\Delta R$ has been found to be at the second
order of electroweak penguin amplitude in the precision of order
$10^{-3}$ and it is sensitive to new physics associating with
electroweak penguin. Of particular, the ratio difference can well
be evaluated in the standard model with isospin and SU(3)
relations. Its numerical values $\Delta R^{SM} \simeq 0.01\sim
0.02$ is found to be much smaller than the current experimental
data $\Delta R^{exp} = 0.18 \pm 0.10$, which strongly indicates a
signal of new physics in the electroweak penguin sector. The
possible new physics effects requires an additional electroweak
penguin contribution with an enhanced amplitude and/or a large CP
phase. In $B\to \pi \pi$ decays, we have demonstrated that the
tree and color-suppressed tree amplitudes $\bar{T}$ and $\bar{C}$
as well as their relative strong phase can independently be
extracted from their three CP-averaging branching ratios, which
are almost not affected by the enhanced electroweak penguin
contributions. The resulting large ratio $|\bar{C}/\bar{T}| \simeq
0.72$ has strongly indicated an enhanced color-suppressed tree
contributions. It has also been shown that the puzzle of a very
large color-suppressed tree amplitude $|C'/T'|\sim 2$ in $B\to \pi
K$ decays may originate from the same reason as the large ratio
difference $\Delta R^{exp}/\Delta R^{SM} > 9.0\pm 5.0$, thus the
puzzle can be solved simultaneously by the same new electroweak
penguin contributions. The large CP violation
$A_{CP}(\pi^0\bar{K}^0) \sim 0.69$  and  $A_{CP}(\pi^0\bar{K}^-)
\sim 0.56$ have also been found to be solved simultaneously with
the new electroweak penguin contributions.

The mixing induced CP violation in $B\to \pi K, \pi\pi$ decays has
also been checked. It has been found that a very large $|C'/P'|$ is
helpful to explain $S_{\pi K_S}$ puzzle, but inconsistent with the
direct CP-violating measurements. And Case III's results are also
consistent with the latest experimental results in $1.5\sigma$ error
level. More accurate measurements on the ratio difference $\Delta R$
and direct/mixing-induced CP violation in $B\to \pi \pi, \pi K$
decays are very important for probing the signal of new physics in
the electroweak penguin sector and testing the isospin and SU(3)
symmetries in the standard model.

\acknowledgments

\label{ACK}

This work was supported in part by the National Science Foundation
of China (NSFC) under the grant 10475105, 10491306, and the
Project of Knowledge Innovation Program (PKIP) of Chinese Academy
of Sciences.

\end{document}